\documentclass[12pt,preprint]{aastex}

\slugcomment{Submitted to ApJ, Feb. 2, 2005, accepted March 23, 2005}

\shorttitle{Transit of Artificial Objects}
\shortauthors{L.F.A. Arnold}

\begin{document}


\title{Transit Lightcurve Signatures of Artificial Objects}


\author{Luc F. A. Arnold}
\affil{Observatoire de Haute-Provence CNRS 04870 Saint-Michel-l'Observatoire, France.}
\email{arnold@obs-hp.fr}

\begin{abstract}
The forthcoming space missions, able to detect Earth-like planets by the transit method, will \emph{a fortiori} also be able to detect the transit 
   of artificial planet-size objects. Multiple artificial objects would produce lightcurves
   easily distinguishable from natural transits. If only one artificial object transits, detecting its artificial nature becomes more difficult. 
   We discuss the case of three different objects (triangle, 2-screen, louver-like 6-screen) and show that they have a transit lightcurve 
   distinguishable from the transit of natural planets, either spherical or oblate, although an ambiguity with the transit of a ringed planet 
   exists in some cases. We show that transits, especially in the case of multiple artificial objects, could be used for the emission of 
   attention-getting signals, with a sky coverage comparable to the laser pulse method. The large number of expected planets (several hundreds) 
   to be discovered by the transit method by next space missions will allow to test these ideas.
\end{abstract}


\section{Introduction}

Current Search for Extraterrestrial Intelligence (SETI) programs concentrate on the search for radio or optical laser pulses emissions \citep{tarter2001}. 
We propose here an alternative approach for a new SETI: considering that artificial planet-size 
bodies may exist around other stars, and that such objects always transit in front of their parent star for a given remote observer, we may thus have an opportunity to detect and even characterize them by the transit method, assuming these transits are distinguishable from a simple planetary transit.
These objects could be planet-size structures built by advanced civilizations, like very lightweight solar sails or giant very low density structures maybe specially built for the purpose of interstellar communication by transit. The mass involved in these structures would be a negligible part of the mass of the mother planet. Such objects would remain much smaller than Dyson shells \citep{dyson1960,bradbury2001}, hypothetical artificial structures bigger than the star itself.

Can the artificial nature of a planet-size body be detected ? 
If only one spherical body transits, there is, strictly speaking, an ambiguity on its nature - natural or artificial -, although spectroscopy (or 
high angular resolution imaging in the future) can give significant insights to answer this question, like for HD209458b \citep{vidalmajar_et_al2003}. 
But we know that a planetary transit lightcurve contains fine features due to the object shape, as soon as a $10^{-4}$ or better accuracy is reached for a 
Jupiter-size object in front of a Solar-like star. These features reveal planet oblateness \citep{baf03}, double planets 
\citep{sarto99} or ringed planets - moreover oblate too - \citep{baf04}.
Although the sphere is the equilibrium shape preferred for massive and planet-size bodies to adapt to their own gravity, one can consider non-spherical
bodies, especially if they are small and lightweight, and transiting in front of a dwarf star to produce a detectable signal. 
Non-spherical artificial objects - like triangles or more exotic shape - have each a specific transit lightcurve, as we show in this paper. 

We also must consider the case of multiple objects: A remarkable lightcurve would be created by free-flyers transiting 
their star successively in a distinguishable manner. At each period, we would observe a series of transits whose number and timing 
would claim its artificial nature and will of communication. 

We will be able to pay attention to these ideas with coming space missions Corot \citep{bor03} and Kepler \citep{borucki_et_al2003}
that will survey up to $6.10^4$ and $10^5$ stars respectively, both reaching a photometric accuracy of $\approx10^{-4}h^{-1/2}$ 
where $h$ is the per-point integration time in hours. 
The Kepler mission expects the detection of hundreds of planets \citep{borucki_et_al2003}. 

Tarter (2001) argued that an advanced technology trying to attract the attention of an emerging technology, such as we are, might do so by producing 
signals that will be detected within the course of normal astronomical exploration. The on-going planetary transit search put us exactly in the position of that
emerging technology civilization.

This paper discusses several possible suspect transits (Section \ref{analysis}). Subsection \ref{simulation} describes how we simulate transit
lightcurves. In Subsection \ref{single}, we point out the difficulty
to detect a compact artificial object from its transit lightcurve shape, while other objects lightcurves could display clearer features usefull
as attention-getting signals. 
Subsection \ref{multiple} presents the possibility of multiple transits of artificial objects, which we consider as more clear attention-getting signals.
In Section \ref{discussion}, we first discuss the sky coverage of the transit emitted signal, comparing it to the laser pulse method (Subsections \ref{direction} and  \ref{comparison}).
We finally briefly present the implications that would rise from null or positive artificial transit detection by transit search programs (Subsection \ref{implications}). 

\section{Analysis of possible suspect transits}
\label{analysis}
\subsection{Simulated transits}
\label{simulation}
The apparent stellar flux during the transit is evaluated from a simulated image of the star in which pixels are zeroed by the transiting object. The flux is normalized to the out-of-transit stellar flux. To have properly continuous stellar flux or planet blocking flux functions (i.e. without steps due to pixelization when planet or star radii change), limb pixels are accurately interpolated. Stellar diameter is $\approx800$ pixels and to save computer time, some integrals are computed only over a strip containing
the transit rather than over the whole stellar disk. The transiting object is assumed on a circular orbit of radius $a=1\ AU$ around HD209458, a 1.15 $R_\odot$ (solar radius) star \citep{brown_et_al2001}. 
The orbit projection on the star practically is close to a straight line although the code treats it as an arc of ellipse. The impact parameter is computed for the transit center. 

The non-planetary object transit is fitted with a planetary transit using a Powell algorithm \citep{press_et_al92} to minimize the square of the difference between both lightcurves. Four parameters in the fitting process are used: The stellar radius  $R_\star$, the stellar limb darkening coefficient $u_1+u_2$ following \citet{brown_et_al2001, baf03, baf04}, the object impact parameter $b=|a\cos(i)|/R_\star$ where $i$ the inclination of the orbit with respect to projected orbit on the sky, and the object radius $r_p$. Two additional parameters, otherwise set to zero, are also considered for the fit when mentioned: Object oblateness $f$ and star limb darkening coefficient $u_1-u_2$, the latter being weakly constrained \citep{brown_et_al2001}. One hundred sampling points were used for the fit, corresponding to a sampling time of the order of $5\ min$.
\subsection{A single object}
\label{single}
We now analyze the transit lightcurve signature of three single non-planetary objects. The first (a triangle) has a simple and compact shape,
while second and third (louver-like) have more elongated and complex structure. All have the same cross-section, equal to a $r_p=1.16$ Jupiter radius ($R_{Jupiter}$) planet.

\begin{figure}[h]
   \centering
   \plotone{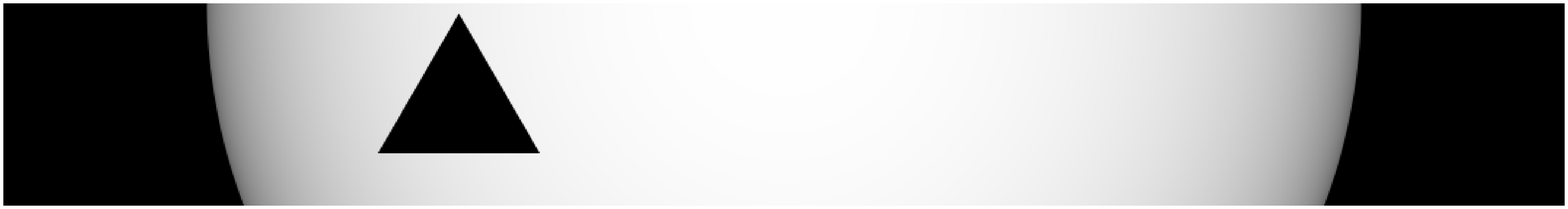}
   \plotone{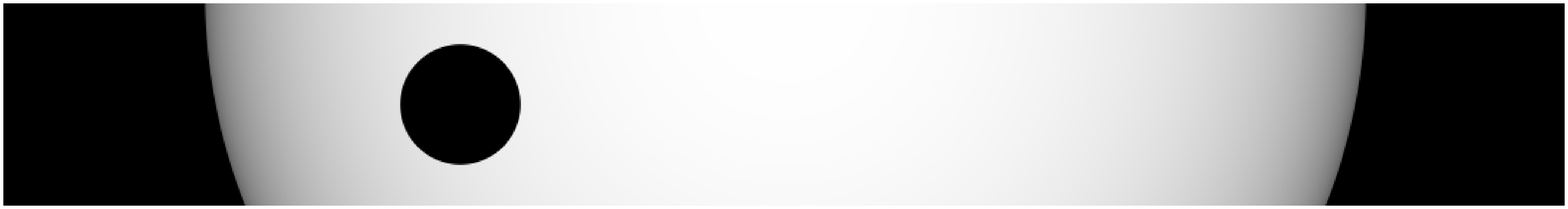}
   \caption{Transiting objects: A triangular equilateral object (upper strip) and the best-fit spherical planet and star (lower strip, same scale as upper strip).
   The star model for the triangle transit is HD209458 with limb darkening coefficients $u_1+u_2=0.64$ and $u_1-u_2=-0.055$ \citep{brown_et_al2001}. The triangle edge length is $0.280$ stellar radius. The object impact parameter is $b=0.176$ (transit center).  The best-fit sphere has an impact parameter of $b=0.19$ and a radius of $r_p=1.16\  R_{Jupiter}$. Best-fit star has $u_1+u_2=0.66$, with $u_1-u_2$ set to zero, and a non-significant radius increase of $0.5\%$. Fitting object oblateness $f$, either with zero or $90^{\circ}$ obliquity to maintain lightcurve symmetry, converges to solutions not significantly different from the case $f=0$.}
   \label{image_transit_tri_up}
\end{figure}

The equilateral triangle is the regular convex polygon that differs most from the (planetary) circle. It is thus able to produce \textit{a priori} a transit lightcurve different from a spherical body transit. 
We consider here an equilateral triangle with one side parallel to the transit direction (Fig. \ref{image_transit_tri_up}). 
This configuration generates a symmetrical lightcurve that eliminates the possibility of transit of a non-zero obliquity oblate (or/and ringed) 
object which is known to produce an asymmetric lightcurve \citep{baf03, baf04}.

\begin{figure}[h]
   \centering
   \includegraphics[width=8.75cm,height=6cm]{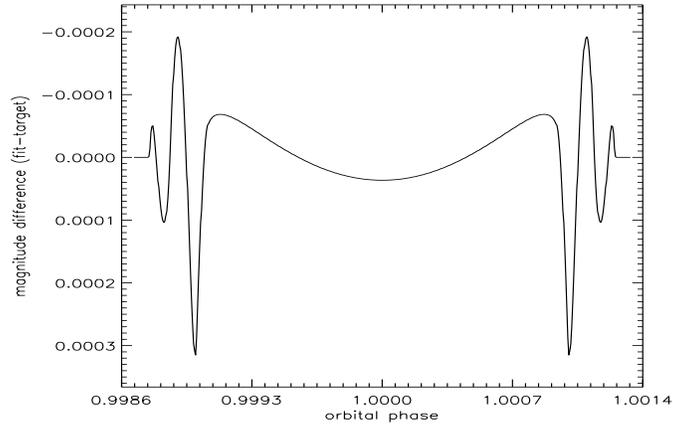}
   \caption{Magnitude difference between the transit of the triangular object and the best-fit sphere shown Fig.\ref{image_transit_tri_up}.}
   \label{residual_transit_tri_up}
\end{figure}

Fig.\ref{residual_transit_tri_up} shows that the triangle transit lightcurve cannot exactly be fitted by a spherical nor an oblate body with zero or $90^{\circ}$ obliquity, and that the fit residuals are symmetrical suggesting a non-spherical zero-obliquity symmetrical transiting body. Our fit gives residuals are of the order of $10^{-4}$, within space missions like Corot or Kepler photometric capabilities of $\approx 10^{-4}h^{-1/2}$  (probably several transits should be cumulated to reach a sufficient accuracy). 

To have a stronger signal, one can imagine these transiting objects in orbit around smaller (dwarf) stars.
Moreover, as we point out in the Introduction, arbitrary shapes rather than spherical are more likely if the size of the object is smaller. 
Therefore, in order to have a detectable signature anyway with a small object, dwarf stars should be preferred by communicative civilizations. 
It must be noted that the curve Fig.\ref{residual_transit_tri_up} resembles a lot to the curve of a ringed planet transit as shown for example in Fig. 1 in \citet{baf04}.
This fitting degeneracy suggests that distinguishing a triangular object from a ringed planet would require a higher accuracy in the transit record
than is mentioned here or in \citet{baf04} to identify planetary rings. 

\begin{figure}[h]
   \centering
   \includegraphics[width=8.75cm,height=6cm]{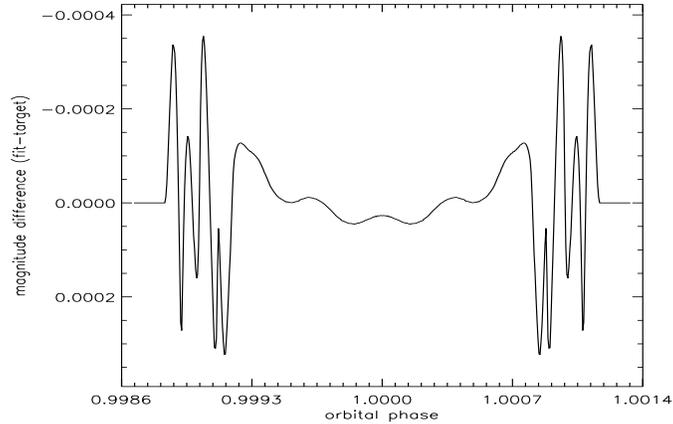}
   \caption{Magnitude difference between the transit of a rotating triangular object (same as shown Fig.\ref{image_transit_tri_up}) and the best-fit spherical planet and star. The triangle makes seven turns on itself during the transit of HD209458 at $b=0.5$. The fit gives a transiting sphere of  $1.17\  R_{Jupiter}$ at $b=0.51$ and a star with $u_1+u_2=0.67$, $u_1-u_2=0$ and $R_\star$ increased by $1\%$. Here, the curve is symmetric because the rotating object is in a symmetric position at transit center with respect to object orbital plane. If it would not be the case, then the curve would be asymmetric.}
   \label{residual_transit_rot}
\end{figure}

If the non-spherical object rotates, the residual
lightcurve will show an additional modulation due to object rotation in front of the background radiance gradient, i.e. the limb darkening of the star  \citep{schneider2004}. This is verified in Fig.\ref{residual_transit_rot}. It shows the magnitude residuals from the lightcurve of a triangle making 7 turns on itself perpendicular to the observer direction after subtraction of the best-fit sphere. At an impact parameter of $b=0.5$ where the background gradient is larger around transit center than at smaller $b$, the modulation outside ingress or egress is of the order of a few $10^{-5}$ magnitude. But object rotation during ingress or egress significantly perturbs the lightcurve which shows in our example several peaks of up to $6\times10^{-4}$ magnitude with abrupt slope variations. 
The magnitude variation ultimately allows the measurement of the object apparent rotation period $p=P/k$, where $P$ is the object rotation period and $k$ the object rotational symmetry order.

\begin{figure}[h]
   \centering
   \plotone{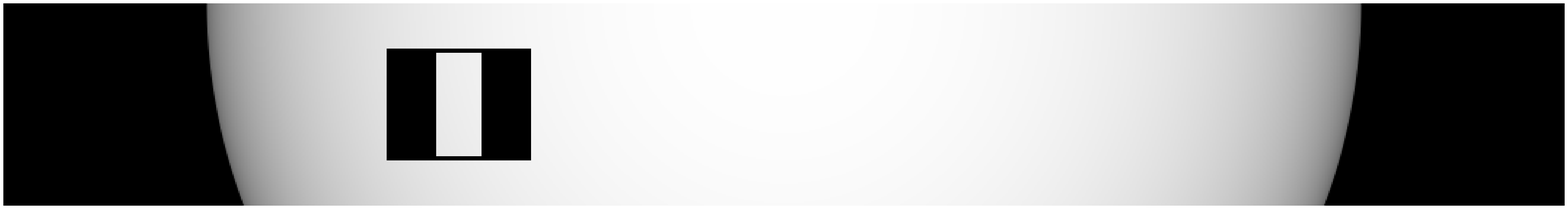}
   \plotone{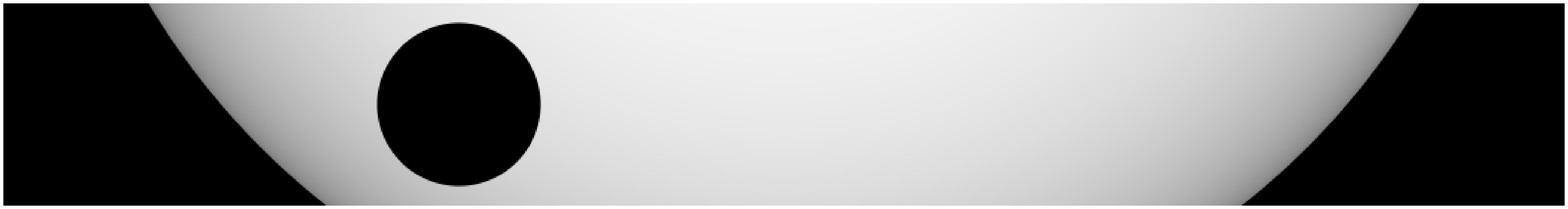}
   \caption{Transiting objects: A 2-screen object (upper strip) and the best-fit spherical planet and star (lower strip, same scale as upper strip).
   The fit gives a transiting sphere of $1.58\  R_{Jupiter}$ at $b=0.51$ and a star with $u_1+u_2=0.67$, $u_1-u_2=0$ and $R_\star=1.47\ R_\odot$.}
   \label{image_transit_louver2}
\end{figure}

\begin{figure}[h]
   \centering
   \includegraphics[width=8.75cm,height=6cm]{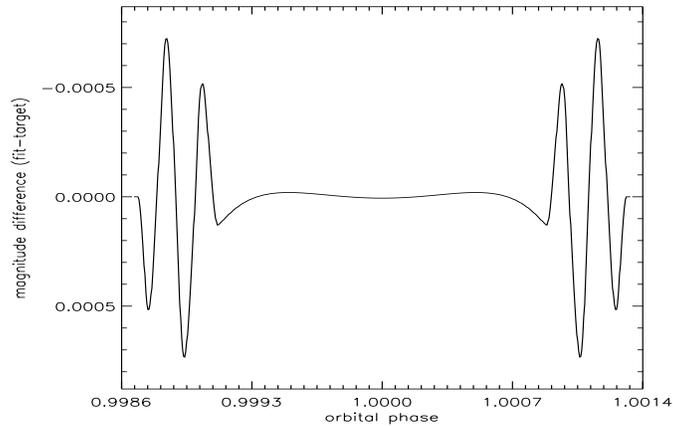}
   \caption{Magnitude difference between the transit of a 2-screen object (Fig.\ref{image_transit_louver2}) and the best-fit spherical planet and star. }
   \label{residual_transit_louver2}
\end{figure}

Let us consider now a two-fold object composed of two symmetrical screens (Fig.\ref{image_transit_louver2}). 
If we neglect the structure linking the screens, the resulting transit can be considered as the sum of two identical lightcurves, slightly phase-shifted, of compact convex bodies. Clearly a sphere would only poorly fit the resulting lightcurve. This is verified on Fig.\ref{residual_transit_louver2} showing residuals three times larger peak-to-peak than for the triangle, although both transiting objects have the same cross-section. For a given impact parameter, the 2-screen object ingress and egress duration is longer than for a compact object of similar cross-section. Therefore the fit by a sphere converges toward a larger impact parameter at which ingress and egress lengthen, and, in order to maintain the overall transit duration and depth, planet and star radii increased by $\approx30\%$. Note that here again, the residuals lighcurve shows an ambiguity with a ringed planet \citep{baf04}.

\begin{figure}[h]
   \centering
   \plotone{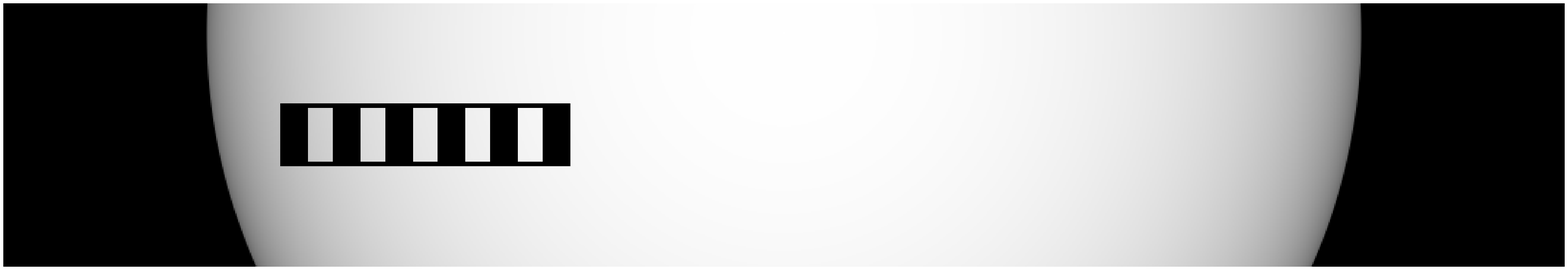}
   \plotone{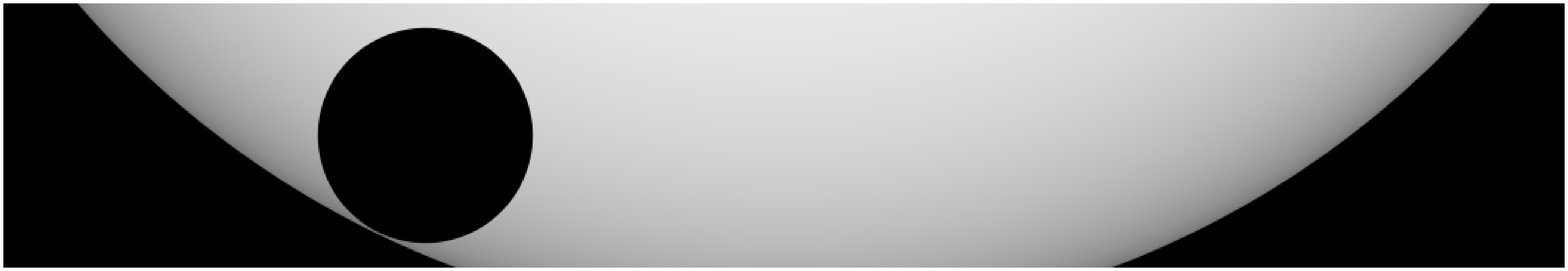}
   \caption{Transiting objects: A louver-like 6-screen object (upper strip) and the best-fit spherical planet and star (lower strip, same scale as upper strip).
   The fit gives a transiting sphere of $2.08\ R_{Jupiter}$ at $b=0.79$ and a star with $u_1+u_2=0.57$, $u_1-u_2=0$ and $R_\star=1.85\ R_\odot$.}
   \label{image_transit_louver6}
\end{figure}

\begin{figure}[h]
   \centering
   \includegraphics[width=8.75cm,height=6cm]{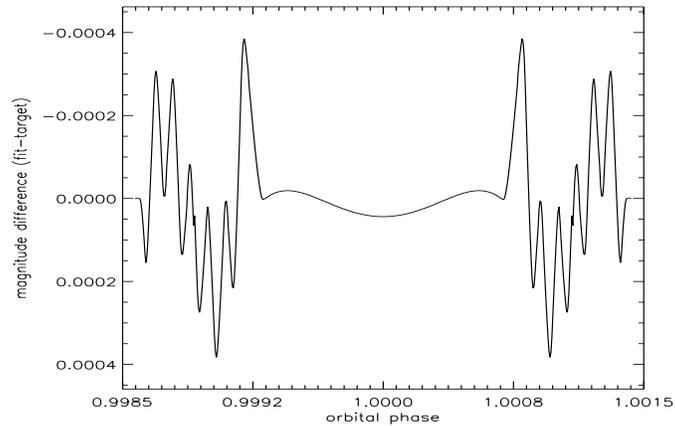}
   \caption{Magnitude difference between the transit of a louver-like 6-screen object (Fig.\ref{image_transit_louver6}) and the best-fit spherical planet and star.  }
   \label{residual_transit_louver6}
\end{figure}

The last example is a louver-like object, an elongated structure composed of six screens (Fig.\ref{image_transit_louver6}). 
The screens produce undulating structures in the ingress and egress transit lightcurve which are
visible in the residuals after best-fit sphere subtraction (Fig.\ref{residual_transit_louver6}). Here again, the object elongated shape induces
a longer ingress and egress than for the previously considered objects: The fit converges toward a larger impact parameter and planet and star radii
increased by $\approx80\%$ and $\approx60\%$ respectively.
The louver practically produces multiple transits. Each screen can indeed be considered as a single objects transiting one after the other, like the two screens
of the previous object. The detectability of the louver is twice better than for the triangle if we consider the peak-to-peak residuals, the cross-section remaining the same. The oscillations during ingress and egress could be considered as an attention-getting signal from a communicative civilization, although the signal is in the $10^{-4}$ range (in our examples) and requires a sampling time of the order of $5\ min$. Detectability can even be higher as we will show in the next Section with multiple transits.

For the record, note also that non-planetary transit could be created by Dyson's sunflowers \citep{dyson2003}, a hypothetical live form spread into space to efficiently collect the energy of a distant star. These structures, if they exist, could be almost circular and quite compact, but also could look like parse or dendritic arrangements.
Nevertheless, large sunflowers are far from the star and transit probability consequently is very low.

\subsection{Multiple objects}
\label{multiple}
Let us now consider a formation of several objects spatially arranged in groups to ingress the star according to a remarkable manner, such as a series of prime numbers, or powers of two. Even more remarkable would be a sudden swap between these two flying formations after several orbits. We consider that such multiple transits would clearly be attention-getting signals and the will of communication would be obvious. The photometric accuracy required to detect $10^{-2}$ or deeper magnitude drops can be reached from the ground. Shallower light drops can be detect from space-based telescope.

A minimal formation would be two objects, either identical or different so the ratio of their cross-sections is remarkable, like integers for example.
The time between their respective ingress could also change, after several identical transits, so their distance would change in a remarkable ratio,
like integers again for example.

\begin{figure}[h]
   \centering
   \includegraphics[width=8.75cm,height=6cm]{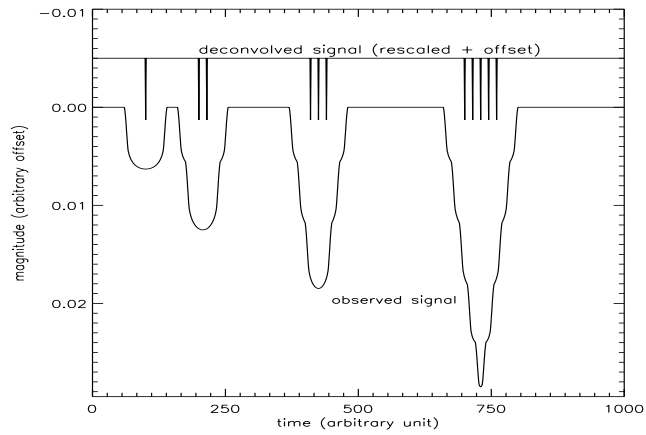}
   \caption{Example of multiple transits generated by eleven objects, grouped in prime numbers (1, 2, 3 and 5 objects). Note that the
   time between transits also increases as a prime numbers series. Each object has a Saturn-like cross-section and transits the star HD209458
   with an impact parameter $b=0.5$. Here, due to objects size and space between them, only 5 objects can be simultaneously
   in front of the star for a given observer. The first transit of a single object allows to deconvolve the transits of multiple objects
   (upper curve).}
   \label{multi_transit}
\end{figure}

If the objects transit individually, i.e. only one object in front of the star at a given moment, then we have a series of identical
drops in the lightcurve, which would look like a binary message (transit/no transit).
If objects are narrower from each other, they transit quasi-simultaneously and the depth of each drop is proportional to the number
of the transiting objects in front of the star at a given time (Fig.\ref{multi_transit}). Assuming all objects are identical, a first single transit gives the key to deconvolve the transit lightcurve of one group, as shown in Fig.\ref{multi_transit} too. The time \textit{between the packets} can be used to introduce information
in the message. In the figure, the time between each transit \textit{inside a packet}
is constant but it can also be modulated to introduce more information inside a packet.

Such flying configurations would not necessarily be
stable on their orbits and would require motors to stabilize the flying formation, except if objects are placed on $L_4$ and $L_5$ Lagrange points
respective to a more massive body, i.e. the mother planet for example.
Due to the relative size of an object with respect to the star, only a limited number of objects can be in front of the star at a given time (five
in our example), but this number can be larger in the case of Earth-size objects, giving more degrees of freedom for informations transmission, with the drawback
of requiring a better photometric sensitivity to be seen.

Compared to a single transit of an artificial (triangular for example as discussed in the previous section), a relaxed photometric accuracy is required to detect the artificial nature of the multiple transits. It is likely that a civilization would rather built a series of (small) objects to generate multiple transits
rather than a (large) single non-spherical object for communicative purpose.

\section{Discussion}
\label{discussion}
\subsection{Directionality and sky coverage of the transit signal}
\label{direction}
Is the transit method efficient for stellar communication compared to communication with laser pulses \citep{kingsley2001} ?
Let us consider that one transit corresponds to one bit of information. Considering a Sun-type star, the period of transit signal is
$T=a^{3/2}$.
This signal is also observable over a solid angle of $\Omega=4\pi\ R_\star/a$ from which the transit
is visible during one period (we neglect the object size with respect to $R_\star$).
We define the mean spatial data rate $b_t$ by the number $n$ of bits emitted over a given solid angle per second of time. The higher this quantity, the higher the
method relevance for the emission of attention-getting signals. Taking into account that one year is $3.2\times10^7s$, we have
\begin{equation}
      b_t={n \Omega\over T}={4\pi\ R_\star\over 3.2\times10^7\ a^{5/2}}\approx\ 2\times10^{-9}\ bit.sr/s
\end{equation}
for an object on an Earth-like orbit ($a=1$). It is increased by a factor of 10 if the object is at $a=0.4$. Another factor of 10 is obtained with $n=10$ transiting objects. The value of $b_t$ could thus reach 
\begin{equation}
      b_t={n \Omega\over T}={10\times 4\pi\ R_\star\over 3.2\times10^7\times 0.4^{5/2}}\approx\ 2\times10^{-7}\ bit.sr/s.
\end{equation}
Precessing objects on a polar orbit would send the transit signal over $4\ \pi\ sr$ but only after several periods. 

For comparison, let us now consider the data rate from laser pulses, $b_p$. 
\citet{kingsley2001} calculated that up to $10^6$ solar-type stars within a radius of $300\ pc$ could be reached by $1\ ns$ laser pulses of $10^{18}\ W$ peak power. 
The farthest target, on the signal reception side, would yet collect a few thousands of photons per pulse with a 10-m telescope. 
On the signal emission side, a visible laser beam produced by a 10-m telescope has a width of $0.01\ arcsec$, covering a solid angle of $\Omega\approx 2\times10^{-15}\ sr$. Assuming a laser duty cycle $t$ of one second, equivalent to a laser mean power of $10^{9}\ W$ following \citet{kingsley2001}, we have
\begin{equation}
      b_p={n\Omega\over t} \approx 2\times10^{-15}\ bit.sr/s.
\end{equation}

Here $b_p$ is much smaller than $b_t$. Nevertheless, one can argue, assuming that life is concentrated near stars only, that the emission of laser pulses elsewhere
than toward stars is ineffective and useless. Let us assume that an angle of $0.1\ arcsec$ is shot around each star, corresponding to $30\ AU$ for a star at $300\ pc$. 
With a 10-m emitting telescope and a beam width of $0.01\ arcsec$, the $0.1\ arcsec$ patch on the sky will require $100\ s$ of time to be shot with the same laser duty cycle as above. Thus the $10^6$ solar-type stars considered by \citet{kingsley2001} would be equivalent to a complete sky coverage ($4\pi\ st$) and $b_p$ becomes
\begin{equation}
      b_p={n\Omega\over t} \approx {4\pi\over 10^{6}\times10^{2}}\approx 1.3\times 10^{-7} \ bit.sr/s.
\end{equation}

We must note that the narrow angle of $0.1\ arcsec$ considered above assumes that ETI has as excellent knowledge of the proper motion and distance of each of its target, in order to shoot with the laser toward the correct patch of sky that anticipates the star motion during the laser propagating time. Otherwise the narrow laser beacon will miss
its target. Without this astrometric knowledge, ETI would have to shoot in a much larger patch of sky around each target: The statistical distribution of stars proper motion measured with Hipparcos \citep{mignard05} within $300\ pc$ is 0.005 to $0.01\ arcsec/yr$. A star at $300\ pc\approx1000\ ly$ typically will move by 
$\approx5\ arcsec$, which would be the typical patch of sky to be shot around a target star. Again with a laser beam width of $0.01\ arcsec$, 
each target star would require of the order of $2.5\times10^5\ s\approx3\ days$ emitting time, and the method would fall in the $b_p\approx 10^{-13} \ bit.sr/s$ range. 
Of course, a smaller emitting telescope could easily produce a wider beacon but, for a given laser pulse energy, with a smaller photon flux, this flux being proportional to the emitting telescope diameter squared.

The values of $b_p$ and $b_t$ above show that both might reach the $10^{-7} \ bit.sr/s$ level, although, as \citet{kingsley2001} wrote in his analysis, there is a lot of room to 'play' with the numbers here. We consider that the sky coverage of the laser and the transit method have a similar efficiency in terms of data rate over a given solid angle.

\subsection{Other aspects for the comparison of transits with laser beacons}
\label{comparison}
The previous section shows that transits seem reasonably suited for the emission of attention-getting signals, as laser beacons are. We leave it to the reader to 
decide which of a $10^{18}\ W$ peak power laser or an Earth or Jupiter-size screen is easier to build. 

Today world's most powerful lasers, now under construction like the Laser
Megajoule (LMJ) in France\footnote{\url{http://www-lmj.cea.fr/}} \citep{andre1999,cavailler2004} or the National Ignition Facility (NIF) in the US\footnote{\url{http://www.llnl.gov/nif/}} \citep{miller04}, will produce pulses of $1.8\times10^6\ J$ within a few $10^{-9}\ s$, reaching instant power of up to $6\times10^{14}\ W$.
This at least remains 1000 times weaker than the pulses invoked by \citet{kingsley2001}. The LMJ should allow up to 600 shots each year (constrains on components cooling, maintenance like optical alignment, optics aging), including only 30 at full power \citep{andre1999,cavailler2004}. This is far below the $\approx10^7$ shots ($1\ Hz$) required each year to reach the previously discussed $b_p$ values. Thus human technology with LMJ or NIF will allow $b_p$ in the $\approx10^{-13} \ bit.sr/s$ range. 

On the large screens side, the Znamya-2 russian experiment in 1993 successfully deployed a $20\ m$ membrane in space \citep{mcinnes99}. Program future developments had considered structures up to $200\ m$, but was stopped in 1999. It can nevertheless be considered as the very first step 
toward deploying large structures in space, although objects cross-section is yet far ($10^{11}$ to $10^{13}\times$) from Earth- or Jupiter-size object. 
Current human technology clearly has $b_t=0$.

Our definition of $b_p$ and $b_t$ does not take into account the maximum distance at which these communicative methods remain efficient. For transits,
$10^{-2}$ drops are detectable on faint and distant stars, like for example in the case of $I=15.72$ magnitude star OGLE-TR-132 discovered with a $1.3\ m$ telescope \citep{bouchy04,moutou04}. 
But it is unlikely that the OGLE telescope is able to discover transits on significantly fainter stars.
OGLE-TR-132 lies at $\approx 2500\ pc$. A $550\ nm$ $10^{18}\ W$ laser pulse emitted from there with a 10-m telescope would yet results in a $\approx$100-photon bunch in a similar Earth-based telescope. The star itself gives $\approx1000\ photons/nm/s$, meaning that with a time resolution better than $0.1\ s$ and a filter width of $1\ nm$, 
the laser still outshines the star. A $1.3\ m$ telescope would only collect 2 photons per pulse, but detection remains possible at this flux level \citep{baz02} with a cooled photomultiplier and adequate time resolution. Therefore, in very first approximation, and leaving aside the knowledge required on target proper-motion mentioned above, both methods may have quite similar ranges.

Transits of opaque objects are achromatic (although one may imagine some spectroscopic coding) and thus detectable over the entire spectrum with a simple CCD, while the detection of $10^{-9}\ s$ pulses requires a more dedicated focal instrument, like a low-resolution spectrograph equipped with a fast photon-counting camera. These arguments emphasize the relevance of artificial transits for stellar communication with respect to laser pulses.

Assuming a civilization controlling both lasers and planet-size artificial structures, it is likely that if transits
are used for the emission of attention-getting signals, laser pulses would be used for the emission of large amount of data 
\textit{in a given direction, i.e. toward a given target}, because the data rate \textit{per sr} can easily be much larger with a laser 
than with the transit method.

\subsection{Implications of artificial transit search results}
\label{implications}
Let us consider the implications of null results.
Among all known variables stars, about $15\times10^3$ are eclipsing binaries, but none is known to have a lightcurve
that suggests artificial eclipsing objects. No Dyson shells have been identified up to now \citep{bradbury2001}, 
although possible candidates exist, i.e. stars with an infrared excess \citep{conroy2003}. 
These negative observations suggest that civilizations able to build star-size structures simply do not exist. 
But it is possible to propose alternative explanations that do not preclude the existence of these civilizations, following and adapting \citet{baz02} 
arguments, for example by invoking the incompleteness of the search in terms of sensitivity or targets.
The detection of zero artificial transit by the end of Corot and Kepler missions would imply the same conclusions too.

A positive result would, among others, obviously motivate other planetary transit search programs. If coming space missions detect one artificial transit among 1000 
new planetary transits, we reasonably can assume a mean value of one communicative civilization for 1000 planetary systems. If this mean value is true, the Poisson law gives a probability of $37\%$ to detect another communicative civilization within another sample of 1000 new discovered transits.

\section{Conclusions}

We have shown that detecting an artificial object from its transit lightcurve shape requires an excellent photometric accuracy. 
In our examples of transiting objects with Jupiter-like cross-sections, the fit residuals are of the 
order of $10^{-4}$. The residuals are not always distinguishable from the transit of a ringed planet. But during ingress and egress,
louver-like objects produce flux periodic variations that could be attention-getting signals from a communicative civilization.
It is more likely that a communicative civilization would privilege multiple transits inducing at least $10^{-3}$ or deeper light drops, more 
easily identifiable as clear attention-getting signals.   

Transit of artificial objects also could be a mean for interstellar communication \textit{from} Earth in the future. We therefore suggest to future human
generations to have in mind, at the proper time, the potential of Earth-size artificial multiple structures in orbit around our star to produce distinguishable
and \textit{intelligent} transits.

\acknowledgments
The author thanks J. Schneider (Observatoire de Paris-Meudon) for stimulating discussions, L. Koechlin (Observatoire Midi-Pyr\'en\'ees, Toulouse), J.W. Barnes and J.J. Fortney (Dpt of Planetary Sciences, U. of Arizona, Tuscon) for their constructive remarks.

\clearpage

\begin{thebibliography}{}
\bibitem[Andr\'{e}(1999)]{andre1999} Andr\'{e}, M.L. 1999, Fusion Engineering \& Design, 44, 43-49
\bibitem[Barnes \& Fortney(2003)]{baf03} Barnes, J.W., \& Fortney, J.J. 2003, \apj, 588, 545-556
\bibitem[Barnes \& Fortney(2004)]{baf04} Barnes, J.W., \& Fortney, J.J. 2004, \apj, 616, 1193-1203
\bibitem[Blair \& Zadnik(2002)]{baz02} Blair, D.G., \& Zadnik, M.G. 2002, Astrobiology, 2, 305-312
\bibitem[Bord\'{e} et al.(2003)]{bor03} Bord\'{e}, P., Rouan, D., \& L\'{e}ger, A. 2003, \aap, 405, 1137-1144
\bibitem[Borucki et al.(2003)]{borucki_et_al2003} Borucki, W.J., Koch, D., Basri, G., Brown, T., Caldwell, D., Devore, E., Dunham, E., Gautier,
T., Geary, J., Gilliland, R., Gould, A., Howell, S., \& Jenkins, J. 2003,
Proc. Conf. on 'Towards Other Earths: DARWIN/TPF and the Search for Extrasolar Terrestrial Planets', 22-25 April 2003, Heidelberg, Germany,
M. Fridlund, T. Henning, Eds., ESA SP-539, 69-81
\bibitem[Bouchy et al.(2004)]{bouchy04} Bouchy, F., Pont, F., , Santos, N.C., Melo, C., Mayor, M., Queloz, D., \& Udry, S., 2004, A\&A, 421, L13-L16
\bibitem[Bradbury(2001)]{bradbury2001} Bradbury, R.J. 2001, Proc. SPIE Vol. 4273 on 'The Search for Extraterrestrial Intelligence (SETI) in the Optical Spectrum III', Stuart A. Kingsley, Ragbir Bhathal, Eds., 56-62
\bibitem[Brown et al.(2001)]{brown_et_al2001} Brown, T.M., Charbonneau, D., Gilliland, R.L., Noyes, R.W., \& Burrows, A. 2001, \apj, 552, 699-709
\bibitem[Conroy \& Werthimer(2003)]{conroy2003} Conroy, C., \& Werthimer, D. 2003, http://setiathome.ssl.berkeley.edu/$\sim$cconroy/ds.html
\bibitem[Cavailler et al.(2004)]{cavailler2004} Cavailler, C., Fleurot, N., Lonjaret, T., \& Di-Nicola J.M., 2004, Plasma Phys. Control. Fusion, 46, B135-B141
\bibitem[Dyson(1960)]{dyson1960} Dyson, F.J. 1960, Science, 131(3414), 1667-1668
\bibitem[Dyson(2003)]{dyson2003} Dyson, F.J. 2003, Int. J. of Astrobiology, 2(2), 103-110
\bibitem[Kingsley(2001)]{kingsley2001} Kingsley, S.A. 2001, Proc. SPIE Vol. 4273 on 'The Search for Extraterrestrial Intelligence (SETI)
in the Optical Spectrum III', 20-26 January 2001, San Jose, California USA, Stuart A. Kingsley and Ragbir Bhathal, Eds., 72-92
\bibitem[Press et al.(1992)]{press_et_al92} Press, W.H., Teukolsky, S.A., Vettering, W.T., \& Flannery, B.P. 1992, {\it Numerical Recipes in C\/},  
Cambridge University Press, $2^{nd}$ Edition
\bibitem[McInnes(1999)]{mcinnes99} McInnes, C.R. 1999, {\it Solar Sailing: Technology, Dynamics and Mission Applications\/}, Springer-Praxis Series in Space Science and Technology,	1st Edition, ISBN 1-85233-102-X
\bibitem[Mignard(2005)]{mignard05} Mignard, F. 2005, private communication
\bibitem[Miller et al.(2004)]{miller04} Miller, G.H., Moses, E.I., \& Wuest, C.R. 2004, Nuclear Fusion, 44(12), S228-S238
\bibitem[Moutou et al.(2004)]{moutou04} Moutou, C., Pont, F., Bouchy, F., \& Mayor, M. 2004, A\&A, 424, L31-L34
\bibitem[Sartoretti \& Schneider(1999)]{sarto99} Sartoretti, P., \& Schneider, J. 1999, A\&A Suppl., 134, 553-560
\bibitem[Schneider(2004)]{schneider2004} Schneider, J. 2004, private communication
\bibitem[Tarter(2001)]{tarter2001} Tarter, J. 2001, Ann. Rev. of Astronomy and Astrophysics, 39, 511-548
\bibitem[Vidal-Majar et al.(2003)]{vidalmajar_et_al2003} Vidal-Majar, A., Lecavalier des Etangs, A., Desert, J.-M., Ballester, G., Ferlet, R., Hebrard, G., \&
Mayor, M. 2003, Nature, 422, 143


\end{thebibliography}
\end{document}